\documentstyle[prl,aps,multicol,psfig]{revtex}
\tighten
\def \be {\begin{equation}}
\def \ee {\end{equation}}
\def \ba {\begin{eqnarray}}
\def \ea {\end{eqnarray}}
\begin{document}
\draft\preprint{\today}
\title{Fractal Noise in Quantum Ballistic and Diffusive
Lattice Systems}

\author{E. J. Amanatidis, D.E. Katsanos and S.N. Evangelou}
\address
{Department of Physics, University of Ioannina, Ioannina 45110,
Greece} \maketitle
\date{\today}
\begin{abstract}

  We demonstrate fractal noise in the quantum evolution of wave packets
  moving either ballistically or diffusively
  in periodic and quasiperiodic tight-binding lattices, respectively.
  For the ballistic case with various initial superpositions
  we  obtain a space-time self-affine
  fractal $\Psi(x,t)$ which verify the predictions by Berry
  for ``a particle in a box", in addition to quantum revivals.
  For the diffusive case self-similar fractal evolution is also obtained.
  These universal fractal features of quantum theory might be useful
  in the field of quantum information, for creating efficient
  quantum algorithms, and can possibly be detectable in
  scattering from nanostructures.

\end{abstract}

\begin{multicols}{2}
\narrowtext

\section{Introduction}

\par
\medskip
A quantum wave packet $\Psi(x,t)$ evolving from a uniform initial
superposition state $\Psi(x,0)$ displays interesting quantum
interference phenomena which involve self-affine fractal quantum
evolution of the probability densities and quantum revivals. This
was shown by Berry \cite{1} in a study of quantum waves confined
in boxes by Dirichlet boundary conditions. It was pointed out
that the choice of hard wall boundaries introduces quantum
fractal noise in the time evolution of uniform deterministic
systems obeying the Schr$\ddot{o}$dinger's equation, which is
expressed via a self-affine fractal probability density
$|\Psi(x,t)|^{2}$. Moreover, over long periods of time
at integer multiples of the revival
time $T_{r}$, the wave packet
$\Psi(x,t)$ returns and reconstructs in its initial form
\cite{2}. In a wide class of circumstances the quantum revivals
are almost perfect with both space and time periodic
$|\Psi(x,t)|^{2}$.

\par
\medskip
Most studies are concerned with spectra having a quadratic
dependence on the quantum number, such as the simple case of the
infinite potential well $E_{j}=E_{1} j^{2}$, $j=1,2,...$, where
perfect and fractional revivals occur at integer and fractional
multiples of the revival time $T_{r}$, respectively \cite{1}. For
quadratic spectra with wave packets localized in energy space
around a quantum number $\overline{j}$ the classical time
$T_{cl}=\pi \hbar /(E_{1} \overline{j})$ and the much longer
revival time $T_{r}=\pi \hbar /E_{1}$ are the only time scales
\cite{3}. For more complex energy dispersions, such as the
Rydberg states $E_{j}=E_{1}/j^{2}$ $j=1,2,...$ higher order time
scales exist apart from the classical, e.g. 1st order, 2nd
order, etc, recurrences, as shown for the ``return to the origin"
probability in the book by Peres \cite{4}. The classical
revolution time for a wave packet starting around the
$\overline{j}$th orbit is $T_{cl}=\pi\hbar
\overline{j}^{3}/(-E_{1})$, while the 1st order recurrences occur
at integer multiples of the revival time $T_{r}=\pi \hbar
\overline{j}^{4}/(-3E_{1})$ \cite{2,3,4}. The classical time
scale $T_{cl}$ defines the time up to which the system follows a
classical orbit while the revival time scale $T_{r}$ resembles
the Ehrenfest time where the classical Liouville density ceases
to describe the system, since quantum interference makes itself
obvious with gradual spreading and distortion of the wave packet.
The classical and revival times are inversely proportional to
the energy eigenvalue separation and its variation, respectively
\cite{4}. The full quantum limit $t\to \infty$ is approached for
much longer times.

\par
\medskip
The quantum evolution of wave packets which consist of a large
number of simultaneously excited eigenstates is related via the
uncertainty principle to ultrashort laser pulses of femtosecond
duration. In this way one can study the evolution of a coherent
superposition of many quantum states. The quantum evolution of
such superpositions can be realised in the physics of Rydberg
atoms \cite{5}, molecules \cite{6}, semiconductor quantum wells
\cite{7}, mode propagation of Bose-Einstein condensates in
magnetic wave guides \cite{8}, laser prepared states in ionic
traps \cite{9}, etc. On the other hand, from studying squeezed
wave packets with reduced spatial uncertainty, in order to behave
like particles as much as possible, one gains a deep insight into
the semi-classical limit where quantum and classical physics meet
according to the correspondence principle. 
The quantum evolution is qualitatively milder than the classical
one and it remembers the initial state, due to the reversible and
unitary nature of the quantum evolution transformation $\exp(-i
Ht/\hbar)$. For classical evolution the initial state always
converges to a stationary distribution. We show that quantum
memory effects exist and a self-affine
fractal space and time distribution of the probability density
$|\Psi(x,t)|^{2}$.
Quantum revivals also appear in the ``return to the
origin" probability. They might be interesting for fundamental
issues, such as the use of quantum mechanics in computation
\cite{10}. It was shown that space-time interference which
repeats itself at discrete time intervals can create a subatomic
quantum counter \cite{11} and can help towards implementation of
factorization for an integer number into primes exploring
fractional revivals \cite{12} (see
also the $N$-slit interferometer in \cite{13}).

\par
\medskip
Our study may also be relevant to another aspect of quantum
information theory, which is the newborn subject of quantum random
walks (e.g. the Hadamard walk) \cite{14}. A quantum walk is
iterated by combining the usual translation with a rotation in
``coin"-space, where a quantum coin-flip replaces the classical
coin-flip. Of course one should avoid any
measurement at intermediate times \cite{15}. The quantum walk has
an improved behavior over the classical random walk since it is
characterized by ballistic (quadratic dependence of the mean
square displacement in time), instead of the usual diffusive (mean
square displacement linear in time) of the classical evolution.
It is a general feature of one-particle  quantum algorithms to speed up
quadratically when compared to classical dynamics \cite{14}.

\par
\medskip
The quantum evolution is analogous to
the classical phenomenon of geometrical optics known as the
Talbot effect \cite{2}. Talbot inspected a coarsely ruled
diffraction grating illuminated with white light through a
magnifying lens. When the lens held close to the grating, the
rulings appeared in sharp focus, as expected. However, moving the
lens away, the rulings of the grating first appeared blurred, as
expected, but then reappeared in sharp focus at multiples of a
particular distance, known as the Talbot distance. In the Talbot
effect the pattern of intensity is doubly periodic, both across
the grating and as a function of distance from the grating. More
discussion on the Talbot effect and its analogies with the
quantum problem can be found in \cite{2}. Although the quantum
revivals are semiclassical in nature 
quantum interference exists since these semiclassical phenomena appear 
only  close to the quantum limit $ t\to \infty$.

\par
\medskip
In this paper, as an alternative to continuous media we aim to
study quantum evolution of wave packets moving in tight-binding
lattices. A discrete and finite space is important for
simulations on a finite computer with the possibility of mapping
it to a quantum computer. This study is a general scheme for
discetizing space when one deals with linear equations (such as
the quantum Schrodinger's equation), which turns out to be
particularly useful when the geometry plays an important role.
It complies with the necessity to exploit possibilities
for the information space, which is also discrete, opposite to
the continuous classical phase space. This convenient
computational tool for implementing space is favored by
the presence of an underlying lattice in solid state
applications. We discuss two cases: (i) the ballistic motion in
the case of zero site potential where quantum interference from
wave scattering at the hard walls creates a sort of quantum
``randomness" similarly to that of the ``particle in a box", and
(ii) a diffusive case via a special quasi-periodic potential of
critical strength \cite{16}. It must be stressed that quantum
diffusive evolution is very different from classical diffusion 
although the space fractal dimension of the quantum probability 
is the same.
The spectrum is of the $\cos$ type
and the stationary eigenstates are simple
$\sin$ elementary trigonometric functions so that the evolution can
be characterized by an infinity of other time scales apart from
$T_{cl}$, $T_{r}$.
In the ballistic case the motion remains ballistic for
all $t$ since the tight binding lattice has no small $t$
semiclassical limit. 

\par
\medskip
Our results confirm the presence of fractal
fluctuations of the probability distribution during the ballistic
and diffusive quantum evolution, in agreement with Berry \cite{1}.
Moreover, 
for all kinds of quantum noise the fractal noise is described 
by the predicted universal fractal
dimensions $D_{x}$, $D_{t}$ and $D_{x=t}$ of the self-affine
curves. For the ballistic motion the mean square distance
increases proportionally to the square of the elapsed time while 
for quantum diffusion the mean square distance
increases proportionally to time.  The self-affinity seen
in the noisy quantum evolution is familiar 
in non-linear systems. Therefore,
techniques from non-linear  physics can be used to understand
some of the mysteries of quantum mechanics, particularly quantum
evolution in open systems in the presence of 
time-dependent potentials.

\par
\medskip
\section{Ballistic quantum evolution}

\subsection{Space-time structures}

\par
\medskip
We have created space-time structures on the discrete orthonormal
finite lattice space basis ${\{|x\rangle, x=1, 2, ..., N\}}$ for various
continuous times $t$. For the ballistic quantum evolution in a
line the mean-square displacement grows quadratically $\langle
x^{2}\rangle = 2 t^{2}$. It can be expressed via the stationary
eigensolutions of the Hamiltonian $H$ through the evolution
operator $\exp(-iHt/\hbar)$, since $H$ is time-independent. The
static Schrodinger's equation $H|j\rangle=E_{j}|j\rangle$,
$j=1,2,...,N$ corresponding to the matrix $H$, which is
tridiagonal for a one-dimensional lattice with zero diagonal
entries and constant off-diagonal elements, has stationary
eigenvalues, eigenvectors $E_{j}=2\cos{{\frac {j\pi}{N+1}}}$,
$|j\rangle=\sqrt{{\frac {2}{N+1}}} \sum_{x=1}^{N} \sin{{\frac
{{j\pi}{x}}{N+1}}} {\frac {}{}}|x\rangle$. The quantum evolution
of the initial ket $|\Psi(0)\rangle$ can be expressed as
 \ba
|\Psi(t)\rangle & = & e^{-iHt/\hbar}|\Psi(0)\rangle \nonumber \\
   & = & \sum_{j=1}^{N}e^{-iE_{j}t/\hbar}|j\rangle\langle
   j|\Psi(0)\rangle
\ea so that the space-time wave function  reads
\ba
   \Psi(x,t) & = & \langle x |\Psi(t)\rangle \nonumber \\
    & = &  \sum_{j=1}^{N}e^{-iE_{j}t/\hbar} c_{x}^{(j)}
   \langle j|\Psi(0)\rangle ,
\ea with the the amplitude on site $x$ denoted  by the elementary
trigonometric function $ c_{x}^{(j)} = \langle
x|j\rangle=\sqrt{{\frac {2}{N+1}}} \sin{{\frac
{{j\pi}{x}}{N+1}}}$.

%
 \centerline{\psfig{figure=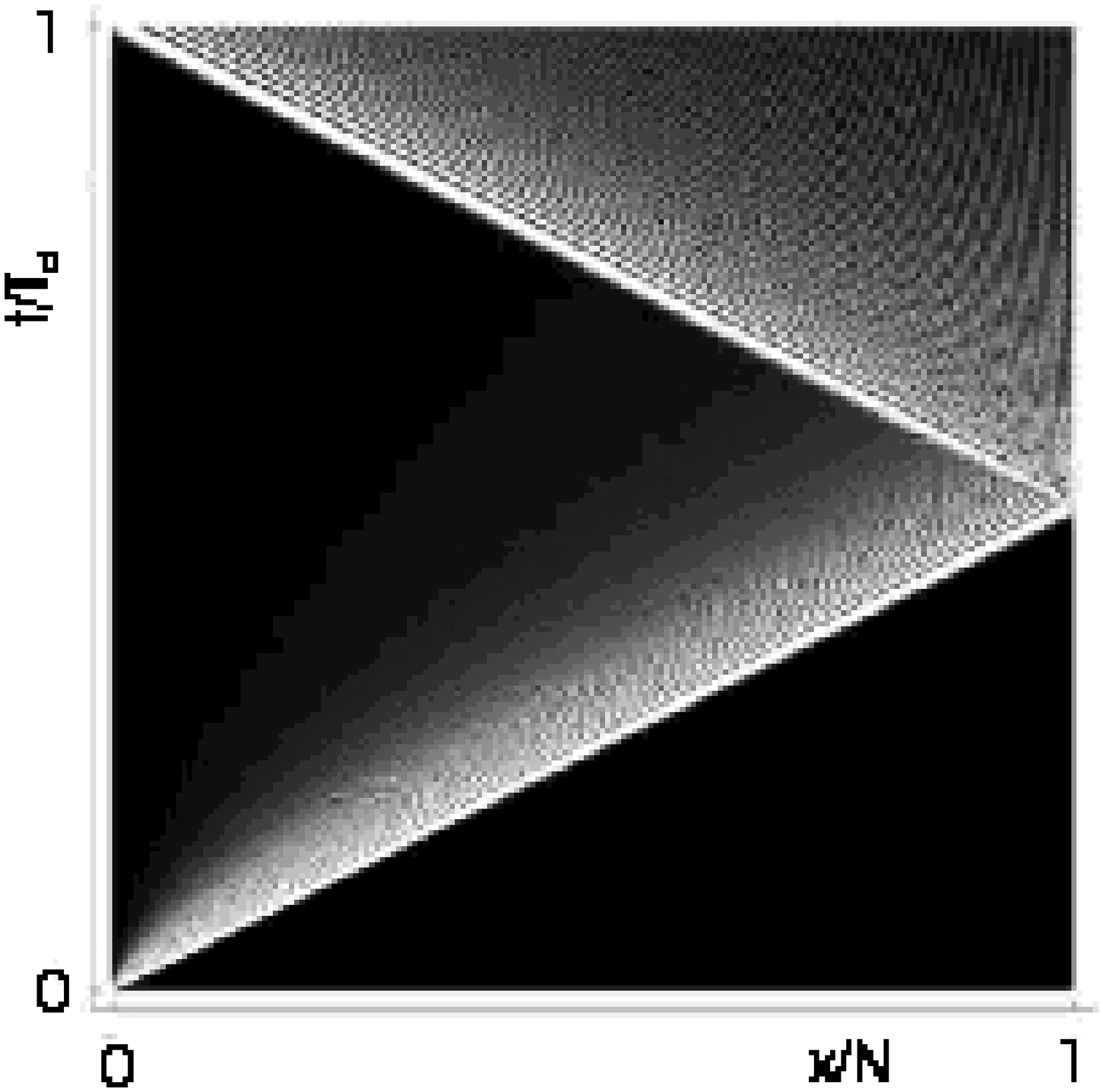,width=8.0cm}}
 \centerline{\psfig{figure=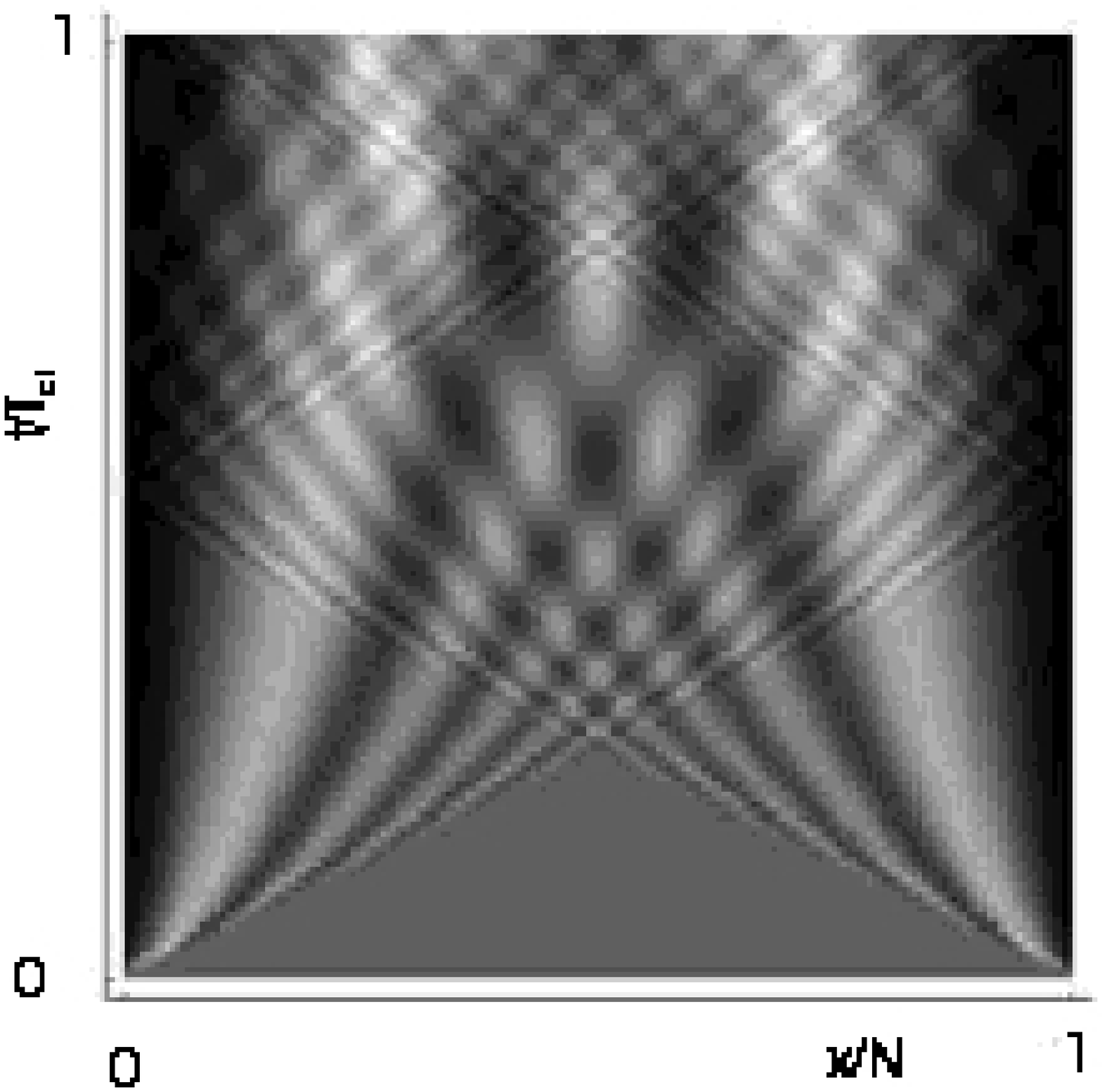,width=8.0cm}}
 \centerline{\psfig{figure=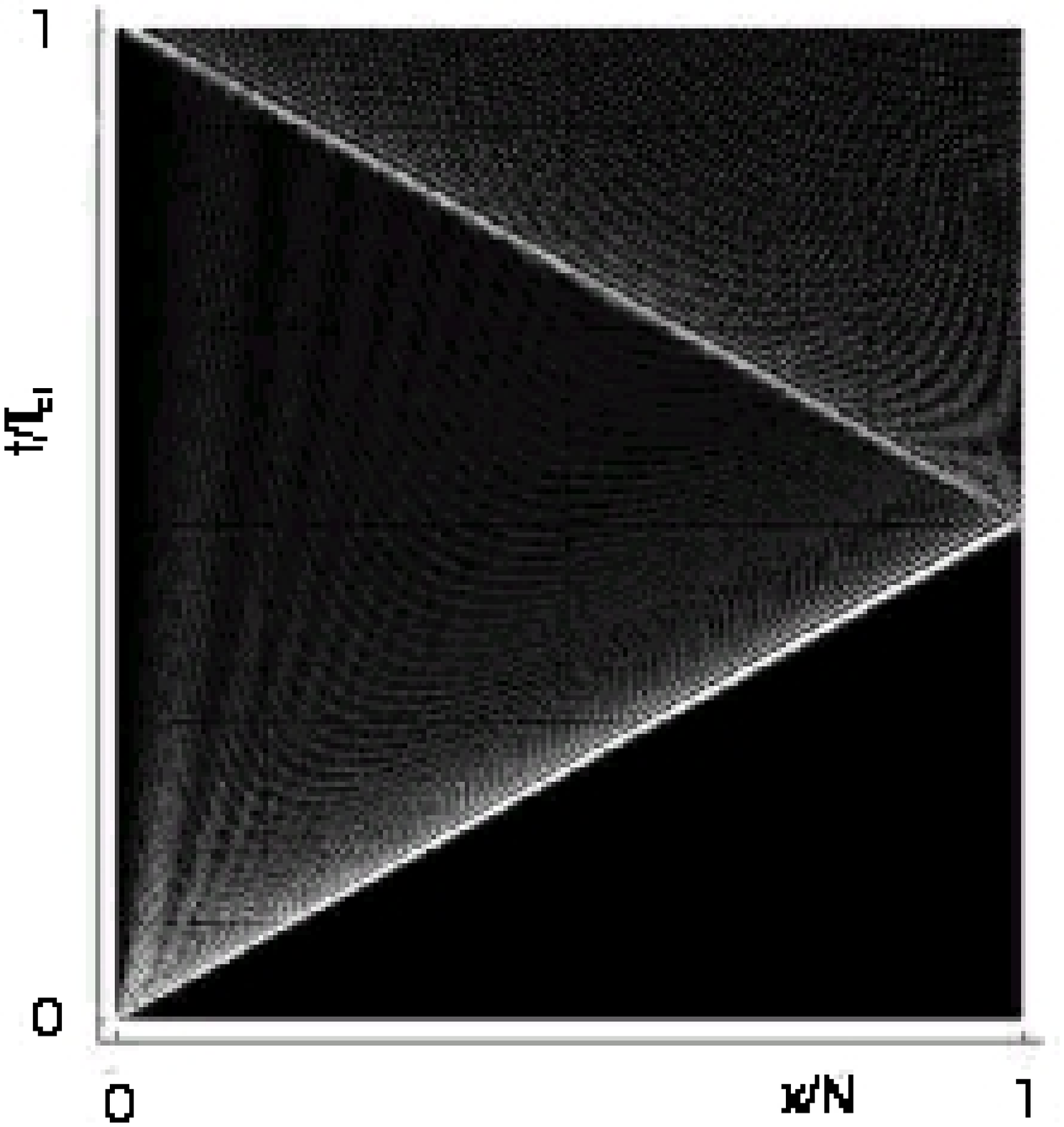,width=8.0cm}}
 \centerline{\psfig{figure=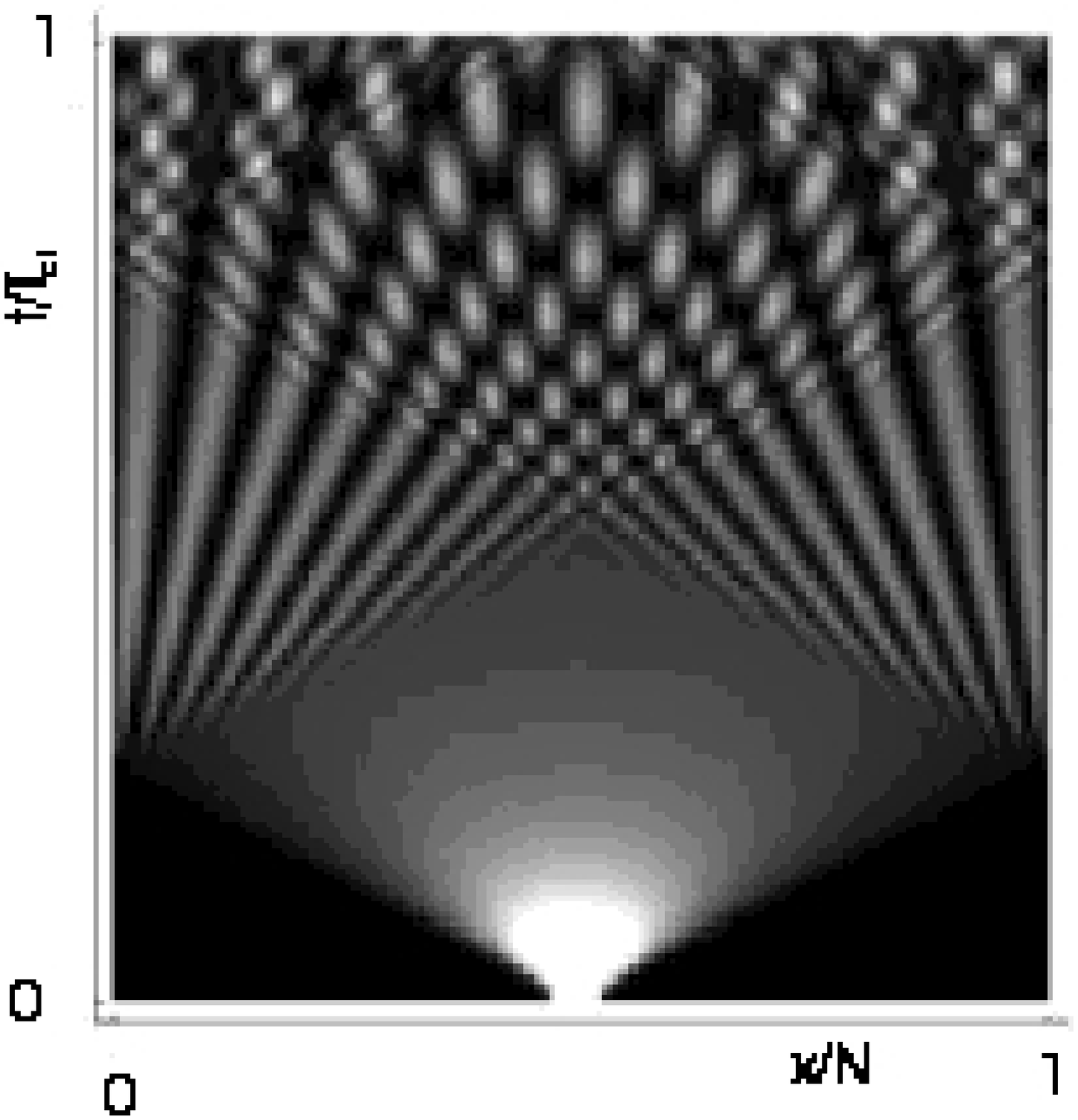,width=8.0cm}}
\par
\medskip
{\footnotesize{{\bf Fig. 1.} The space-time ($x$-$t$)
representation of various initial wave packets moving
ballistically in a chain. The classical time is $T_{cl}\approx \hbar
(N+1)$ and quantum memory effects are seen to develop with
different initial wave packets giving different pictures.
{\bf(a)} For a local initial state $\Psi(x,t=0)=\delta_{x,1}$ in
position $1$ with $N=1000$ we see almost classical motion with
recurrence at $t=1001$. {\bf(b)} For a spatially uniform initial
superposition $\Psi(x,t=0)=1/\sqrt{N}$ in a chain with $N=100$ we
observe interesting interference effects. {\bf(c)} For an initial
uniform superposition of all stationary eigenstates with
$|\Psi(0)\rangle=\sum_{j=1}^{N} a_{j}|j\rangle$,  with
$a_{j}=1/\sqrt{N}$ for any $j$, we have
$\Psi(x,t=0)=\sqrt{2/N(N+1)}\sum_{j=1}^{N} \sin{\frac{j\pi
x}{N+1}}$ amplitude at site $x$ for a chain with $N=1000$. The
result resembles that of (a) where a wave is released at one end
of the chain since in the spatial basis this initial state has
higher amplitude near the end of the chain with diminishing
oscillations. {\bf(d)} For a Gaussian in space initial superposition
$\Psi(x,t=0)=a_{x}$ for $N=100$ interesting interference 
patterns are also seen.}}

\par
\medskip
The ballistic evolution is considered for various initial
superpositions $\Psi(x,0)=\langle x|\Psi(0)\rangle$, chosen from
expansions either in the spatial basis with $|\Psi(0)\rangle=
\sum_{x=1}^{N}a_{x} |x \rangle$, $\sum_{x=1}^{N}|a_{x}|^{2}=1$ 
or in the eigenstate basis with
$|\Psi(0)\rangle= \sum_{j=1}^{N}a_{j} |j \rangle$,
$\sum_{j=1}^{N}|a_{j}|^{2}=1$.
The amplitude $c_{x}^{(j)}$ is real and the transformation from the
spatial $|x\rangle$ to the eigenstate $|j\rangle$ basis or its
inverse from $|j\rangle$ to $|x\rangle$ are the same. In this
case the computations of the space-time wave function $\Psi(x,t)$
at various times are particularly simple by replacing the
eigenvalues $E_{j}$, eigenvector amplitudes $c_{x}^{j}$ in Eq.
(2). For 
initial superposition in the spatial basis the
normalized space-time probability becomes
\ba
  P(x,t)     = |\Psi(x,t)|^{2}
      & = &\left|\sum_{j=1}^{N}
       e^{-iE_{j}t/\hbar}(\sum_{x'=1}^{N}c_{x'}^{(j)*}a_{x'}) c_{x}^{(j)} \right|^2
\ea 
and for superposition in the eigenbasis
the normalized space-time probability is
 \ba
  P(x,t) & = & |\Psi(x,t)|^{2}=\left|
  \sum_{j=1}^{N}e^{-iE_{j}t/\hbar}
         c_{x}^{(j)}a_{j}\right|^{2}.
\ea

\par
\medskip
Firstly, the initial wave was released on a ${\it local}$ state
at one end of the chain with $|\Psi(0)\rangle=|1\rangle
\Rightarrow \Psi(x,0)=\langle x|\Psi(0)\rangle=\delta_{x,1}$
which from Eq. (3) at later times gives the probability \be
  P(x,t)= |\Psi(x,t)|^{2}=
  \left|\sum_{j=1}^{N}e^{-iE_{j}t/\hbar}
  c_{1}^{(j)*}c_{x}^{(j)}\right|^2,
\ee in terms of the eigenvalues $E_{j}$, eigenvectors
$|j\rangle=\sum_{x=1}^{N}c_{x}^{(j)}|x\rangle$. This gives an
almost classical evolution of a point particle seen in Fig. 1(a)
where the particle at time $T_{cl}= \hbar (N+1)$ returns
to its initial position. The quantum revival resembles a
classical particle moving with constant velocity which reflects
at the boundaries at $t=T_{cl}$ and  returns to its original
position $1$.
Our second initial choice was the spatial superposition in space
$|\Psi(0)\rangle=\sum_{x=1}^{N}a_{x}|x\rangle$ with ${\it
uniform}$ $a_{x}=1/{\sqrt{N}}$ for all $x$, which gives from Eq.
(3) the normalized space-time probability  \ba
  P(x,t) =|\Psi(x,t)|^{2}
       & = & {\frac {1}{N}} \left|\sum_{j=1}^{N}
       e^{-iE_{j}t/\hbar}\sum_{x'=1}^{N}c_{x'}^{(j)*}
       c_{x}^{(j)}\right|^2.
\ea
The corresponding space-time structure which arises from the
initial condition of Eq. (6) is shown in Fig. 1(b) where we 
can observe interesting  quantum interference features.
The third choice we have made was not spatial but a ${\it
uniform}$ ${\it expansion}$ ${\it on}$ ${\it all}$ ${\it
stationary}$ ${\it eigenstates}$ $|j\rangle$ choosing the initial
superposition $|\Psi(0)\rangle= \sum_{j=1}^{N}a_{j}|j\rangle$,
with $a_{j}=1/{\sqrt{N}}$ for all $j$. If transformed to the
spatial basis this choice has a maximum at one end of the chain
with decaying oscillations. At later times the normalized
space-time probability is
\ba
  P(x,t) = |\Psi(x,t)|^{2}
       & = & {\frac {1} {N}} \left |
  \sum_{j=1}^{N}e^{-iE_{j}t/\hbar}
         c_{x}^{(j)}\right|^{2}.
\ea
Finally, a ${\it Gaussian}$ spatial initial superposition was
chosen  of the type $|\Psi(0)\rangle=\sum_{x=1}^{N}a_{x}|x\rangle$,
centered in the middle of the chain $N/2$. In this case after
normalization the probability amplitude can be obtained from Eq. (3)
with $|a_{x}|^{2} \propto e^{-(x-N/2)^{2}/(2\sigma^{2})}$,
$\sum_{x=1}^{N}|a_{x}|^{2}=1$, drawn from a Gaussian with
$\sigma^{2}=1$.

\par
\medskip
The obtained space-time structures for the various initial from
the second to the fourth conditions are shown in Fig. 1. The
white denotes high amplitude and the amplitude is lower in the
darker. The spectacular patterns shown are known as quantum
carpets \cite{2} and arise from scattering at the hard walls at
the ends. The quantum intereference is more prominent for the
uniform initial condition (Fig. 1(b)) where the picture resembles
more closely the quantum carpets of Berry \cite{1}. We can also
see quantum recurrences more clearly in
Fig. 1(a), (c) where the initial wavepacket has maximum amplitude
close to the end site $1$.  The
quantum carpets shown in Fig. 1. repeat indefinitely at time
$T_{cl}$, but less and less less accurately each time.

\subsection{Fractal dimensions for
the probability density}

\par
\medskip
We observe the quantum evolution for a uniform initial wave
function along the chain as a function of space freezing the time, 
as a function of time at a single site 
or varying space positions equal to time. The corresponding
pictures are shown in Figure 2 where interesting self-affine
fractal oscillations are seen. For the spatial distribution our
simple computations for the fractal dimension coincide with the
value $D_{x}=1.5$ predicted by Berry. This value coincides 
with the displacement curve of classical random walks which gives
Brownian noise. Quantum ``randomness" described by the
fractal dimension $D_{x}$ arises from the linear Schrodinger's
equation, only due to quantum interference from scattering at the
boundaries. It must be noted these curves are not self-similar
but self-affine fractals. The computed time fractal dimension
also gives the predicted value $D_{t}=1.75$. The $D_{t}$ is
higher than $D_{x}$ as it can be seen form the more noisy
oscillating of the probability density as a function of time t.
We also display a space-time function along the diagonal $x=t$
where the self-affine fluctuations are much smaller with
$D_{x=t}=1.25$.

%
\par
 \centerline{\psfig{figure=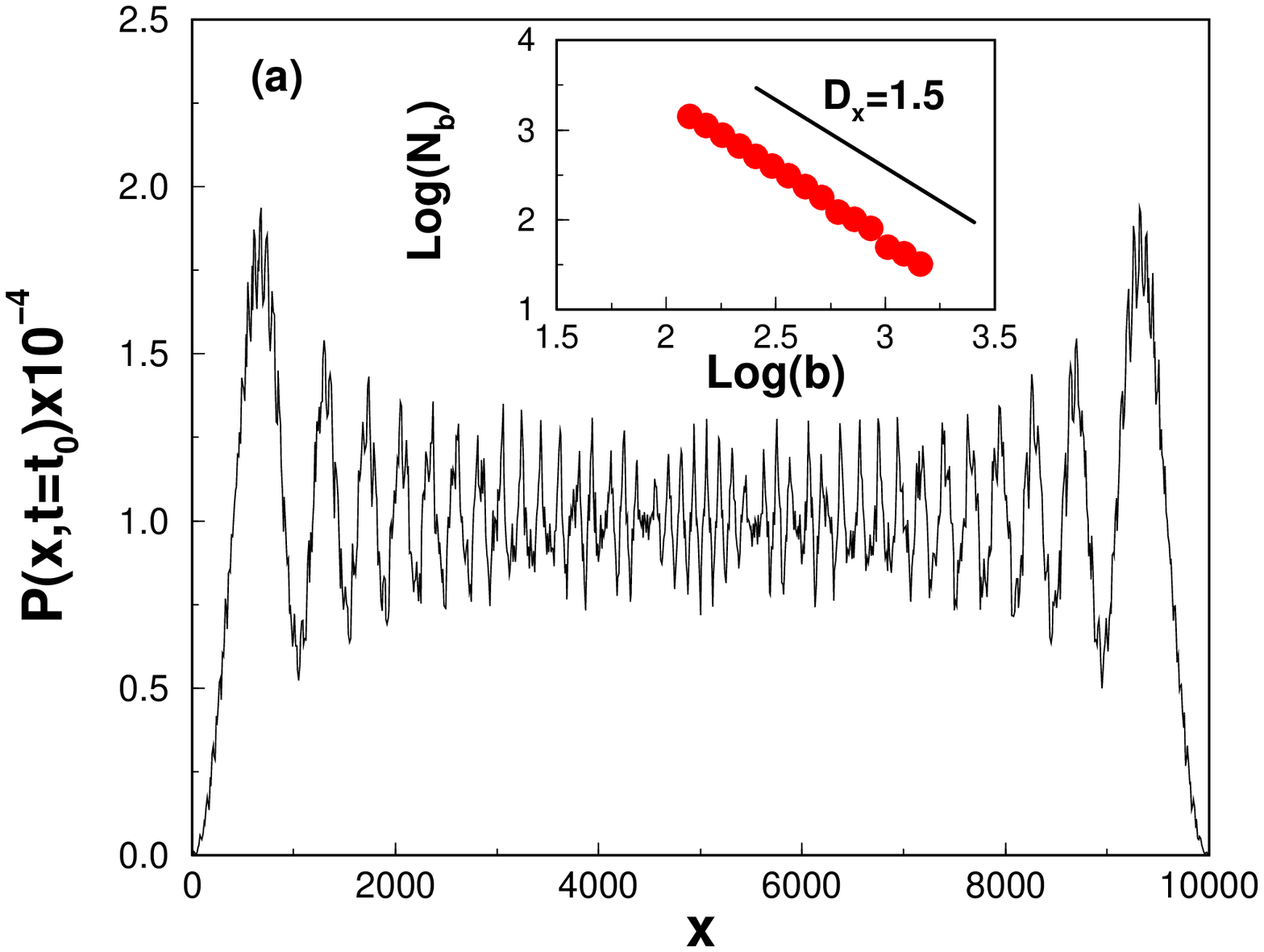,width=8.0cm}}
 \centerline{\psfig{figure=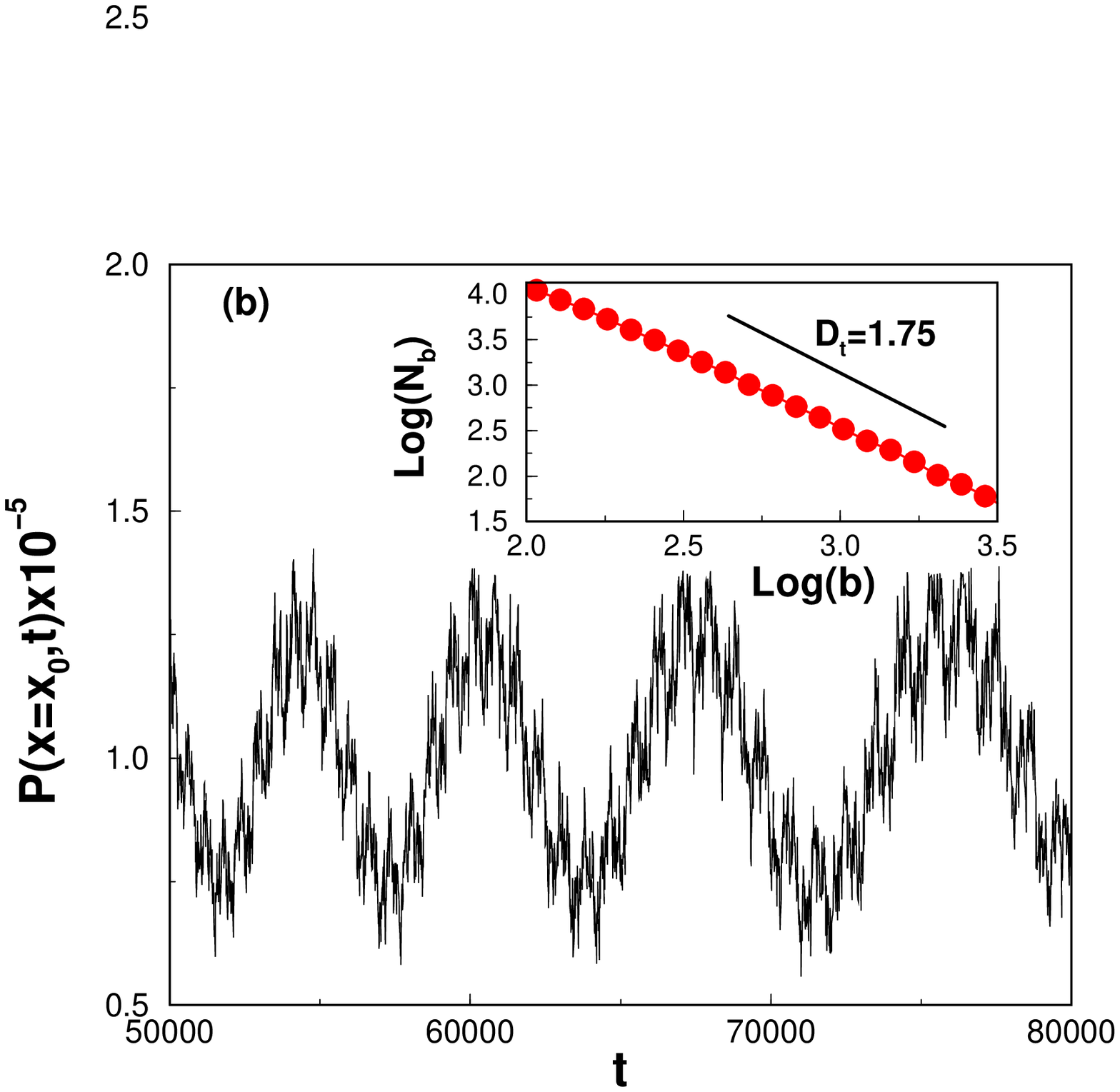,width=8.0cm}}
 \centerline{\psfig{figure=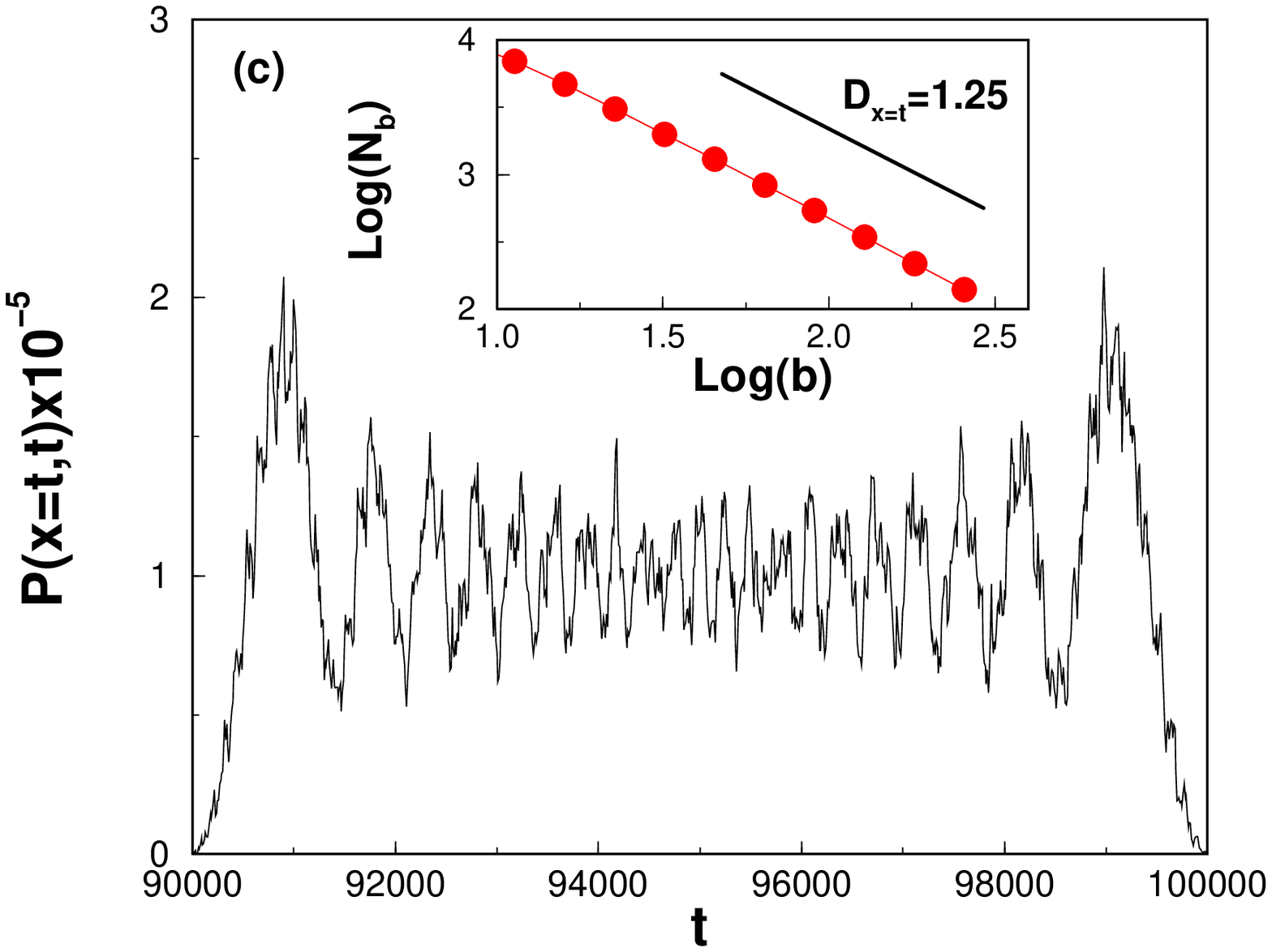,width=8.0cm}}
\par
\medskip
{\footnotesize{{\bf Fig. 2.} Fractal noise
of an initially uniformly distributed in space wave packet 
for a chain of length $N=10000$. 
\bf (a)} The absolute
of the probability amplitude squared $|P(x,t)|^{2}$ as function
of space $x$ at fixed time $t_{0}=100000$.
{\bf (b)} 
A function of time $t$ at the fixed site $x_{0}=5000$. {\bf (c)}
A function of both space and time with $x=t$. The obtained
fractal dimensions are denoted in the log-log insets on the plot.}

\par
\medskip
The capacity (box) dimension of the self-affine $y=y(x)$ curves is
computed by counting boxes of linear size $b . \tau$ along the
x-axis axis and of size $b . a$ along the y-axis. It must be
noted that a similarity dimension cannot be defined 
since for a self-affine curve scaling is different along the two
axes. We have used integer time units so the minimum box width
along the x-axis is taken $\tau =1$ and use a high-resolution
grid for the amplitude along the y-axis. This is required in
order to obtain the ``local" non-integer fractal dimension $D$ of
the self-affine curves, while larger values of $a$ finally
give the well known ``global" fractal dimension $D=1$ \cite{17}.
The selected values of $a$ were of order of $ 10^{-10}$ for both
the space $D_{x}$ and time $D_{t}$ dimensions. For the space-time
fractal $a$ is of the order of $7\times 10^{-8}$. The 
box dimension is defined by the scaling equation $N(b,a)\sim
b^{-D}$, where $N$ is the number of boxes needed to cover the
curve \cite{17}. A linear fit of the scaling equation is
displayed in the log-log plots of the corresponding insets of
Fig.2  which gives $D$ from the slope. As pointed out above the
obtained results for $D_{x}, D_{t}, D_{x=t}$ are in agreement
with the values for the fractal dimensions predicted by Berry
\cite{1}. For the other initial conditions considered we find
similar fractal dimensions although
with errors arising from the
small sizes (see Fig. 3).

\par
\medskip
%
 \centerline{\psfig{figure=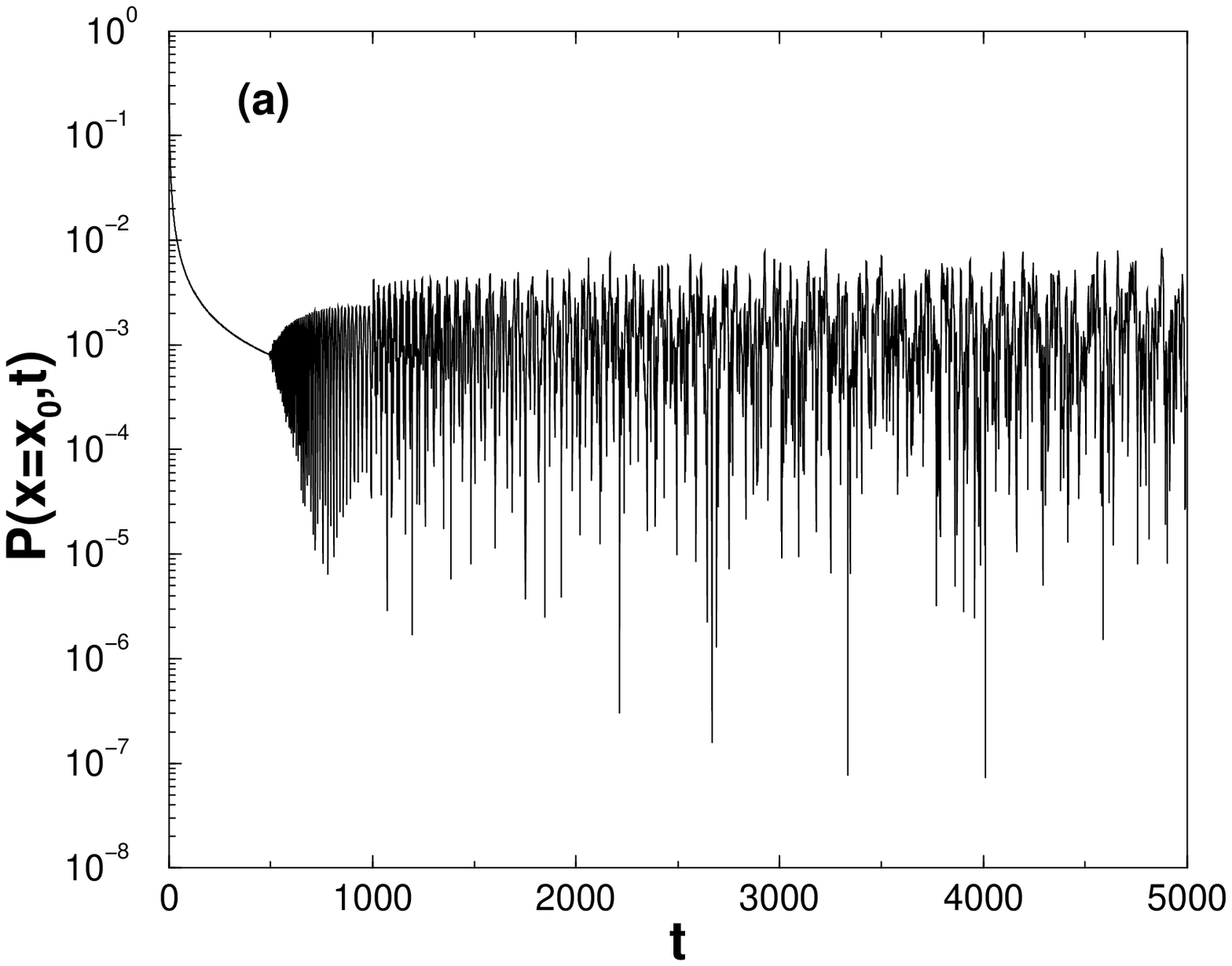,width=8.0cm}}
 \centerline{\psfig{figure=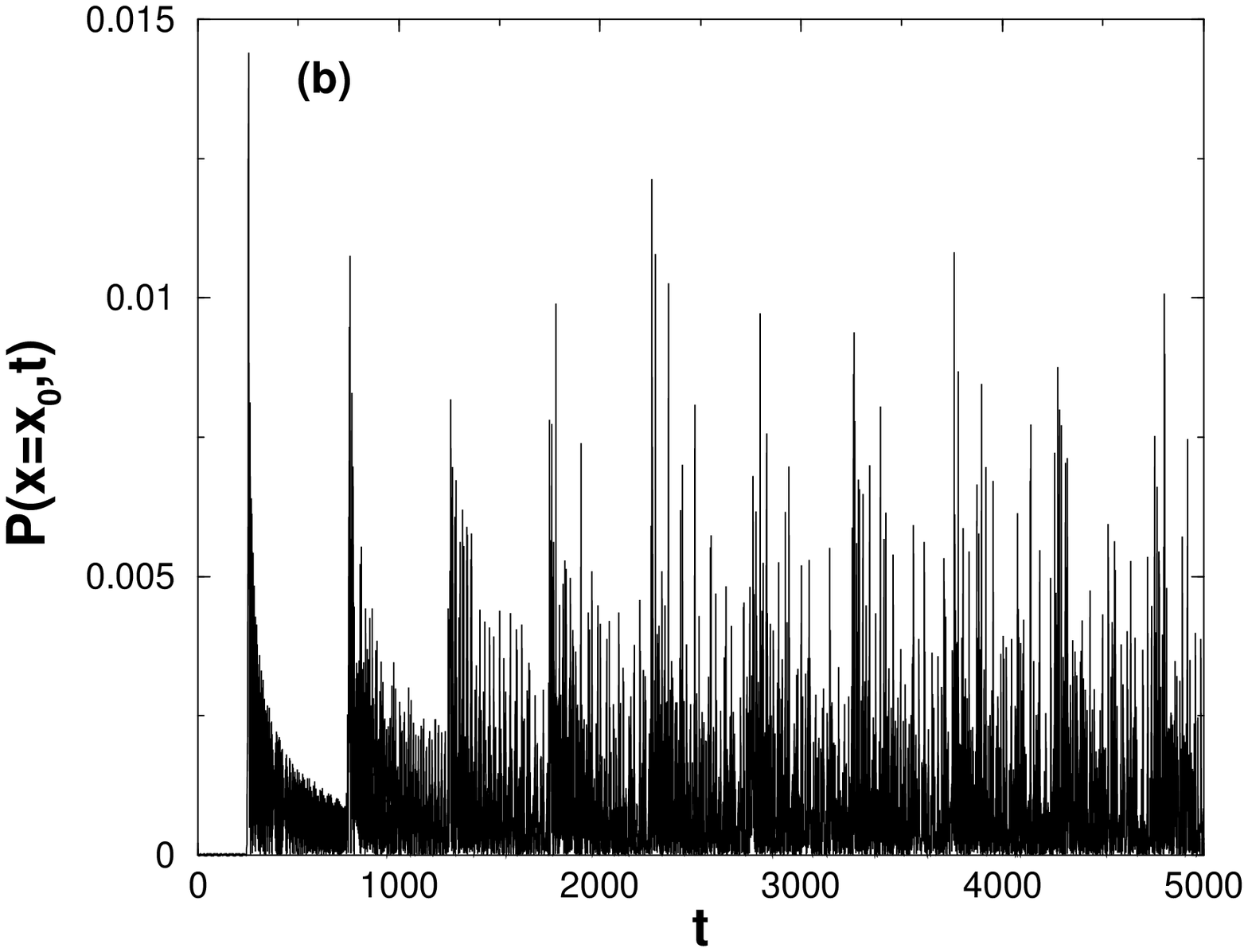,width=8.0cm}}
\par
{\footnotesize{{\bf Fig. 3.} Time fractals for other initial
conditions in a chain of length $N=1000$. 
{\bf (a)}. A Gaussian spatial superposition at a
fixed site $x_{0}=500$ where we
compute $D_{t}\approx 1.80$. {\bf (b)} A uniform initial
superposition in the eigenbasis where we find $D_{t}\approx
1.46$.}}

\subsection{``Return to the origin" probability}

\par
\medskip
The ``survival" or ``return to the origin" probability is defined
from the squared absolute amplitude  overlap \ba P_{0}(t)=|\langle
\Psi(0)|\Psi(t)\rangle|^{2} \ea which can be computed from Eq.
(1). For an initial superposition in the spatial basis
$|\Psi(0)\rangle=\sum_{x=1}^{N}a_{x}|x\rangle$
one obtains \ba P_{0}(t)
        & = & \left|\sum_{j=1}^{N}\sum_{x=1}^{N}\sum_{x'=1}^{N}
       e^{-iE_{j}t/\hbar}a_{x}^{*} a_{x'}
       c_{x'}^{(j)*}c_{x}^{(j)}\right|^2.
 \ea
For an initial superposition in the eigenstate basis
$|\Psi(0)\rangle=\sum_{j=1}^{N}a_{j}|j\rangle$
we have the simpler expression
independent of eigenvectors
 \ba
P_{0}(t)
        & = & \left|\sum_{j=1}^{N}
       |a_{j}|^{2} e^{-iE_{j}t/\hbar} \right |^2
       \nonumber \\
        & = & \sum_{j=1}^{N} \sum_{j'=1}^{N}
       |a_{j}|^{2} |a_{j'}|^{2} e^{-i(E_{j}-E_{j'})t/\hbar}.
\ea

\par
\medskip
%
 \centerline{\psfig{figure=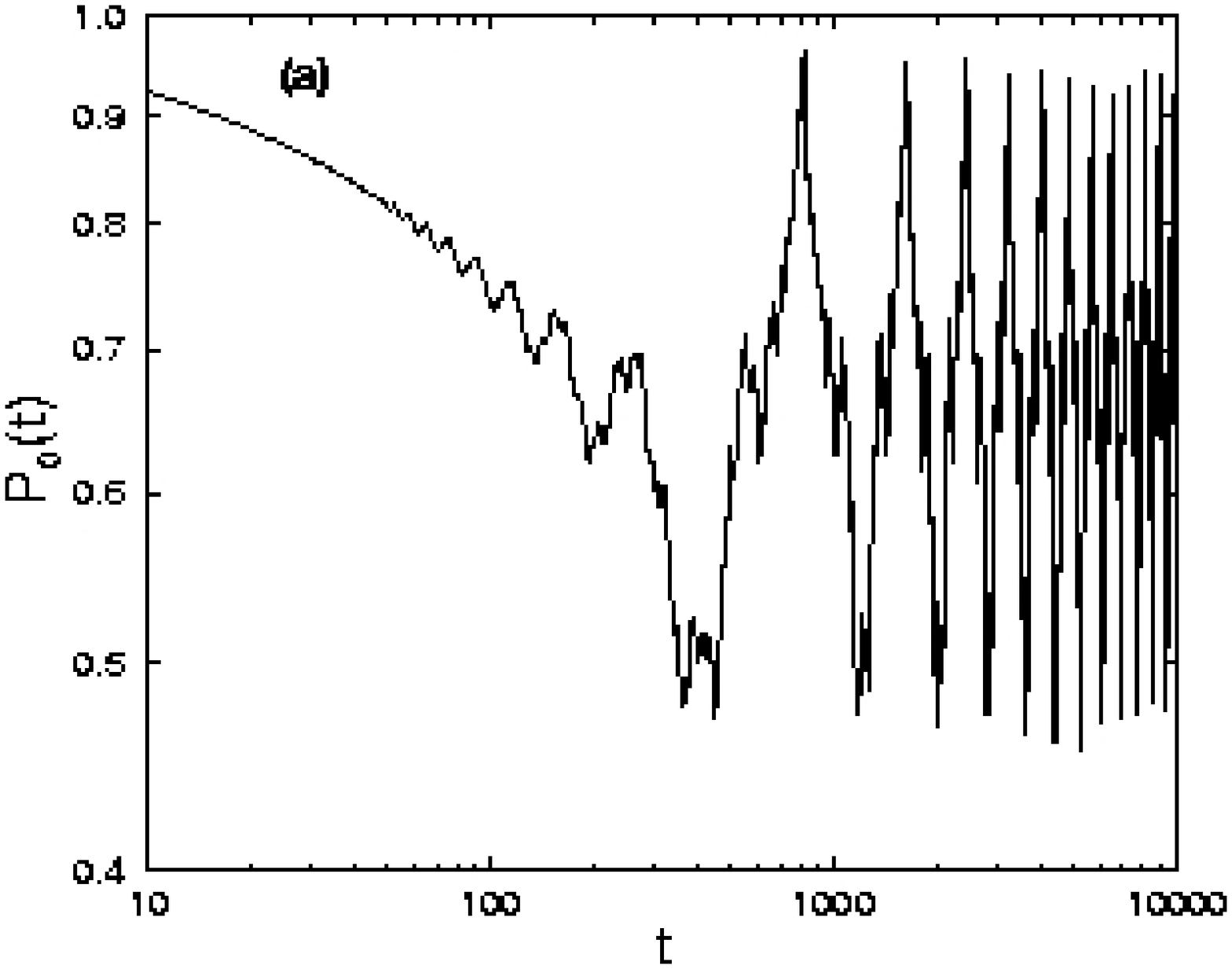,width=8.0cm}}
 \centerline{\psfig{figure=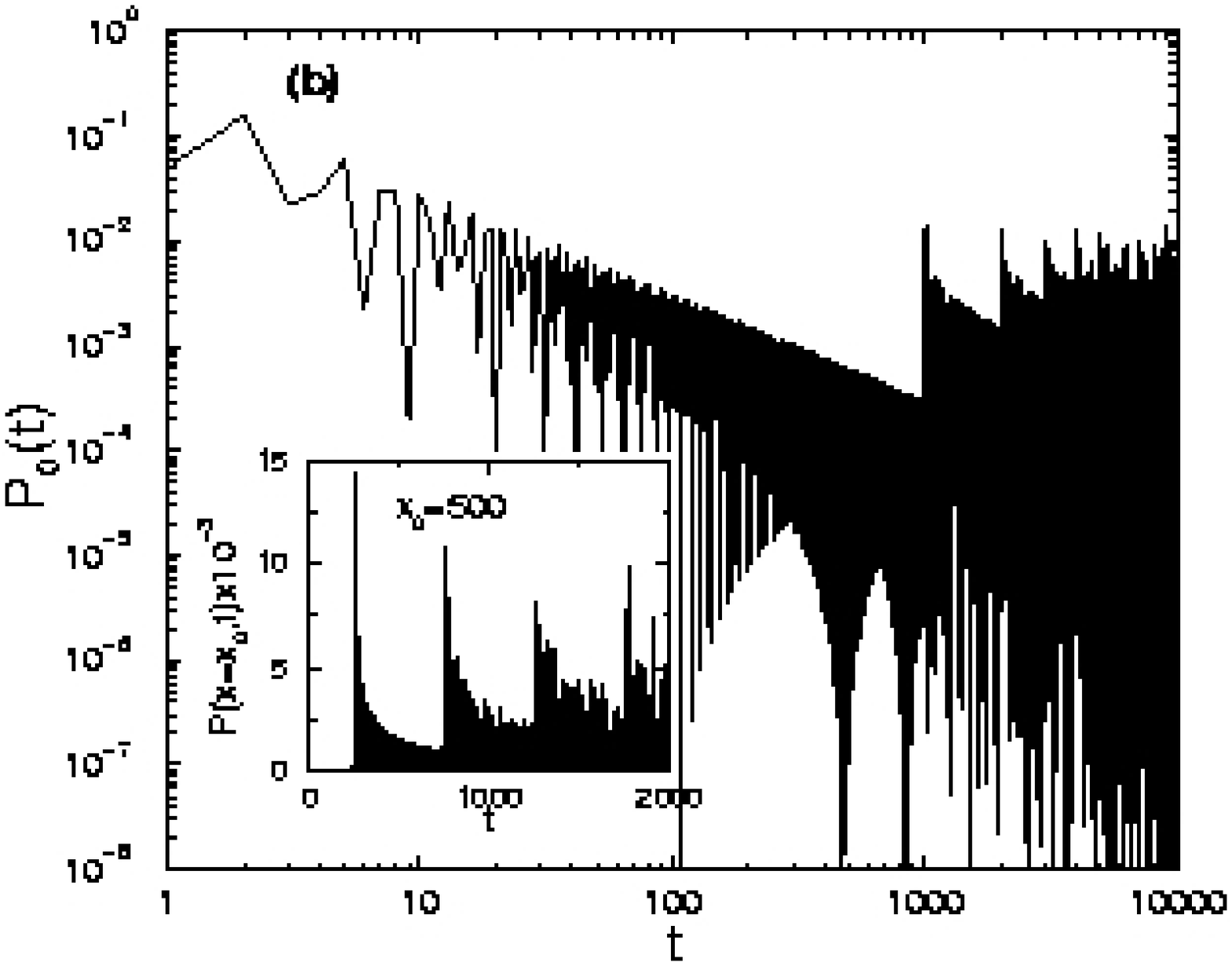,width=8.0cm}}
\par
{\footnotesize{{\bf Fig. 4.} The ``return to the origin"
probability $P_{0}(t)$. {\bf (a)} For an initially uniform
superposition in space and a chain of length $N=100$. The quantum
revivals occur at an estimated period $T_{r}\propto \hbar
(N+1)^{2}\approx 0.08(N+1)^{2}$. {\bf (b)} For an initially
uniform superposition of stationary eigenstates in a chain of
length $N=1000$. In the inset a presentation of the probability
density $P(x,t)$. The quantum revivals which appear with period
$T_{cl}\propto \hbar (N+1)$. }}

\par
\medskip
 The quantum revivals can be seen in Fig.4 where $P_{0}(t)$ is shown
 versus $t$ in a log-log plot. $P_{0}(t)$ is periodic in the early
 stages of evolution which becomes less clear later.
 The long-time overall asymptotic algebraic
 decay $P_{0}(t)\propto {\frac {1} {t}}$ of ballistic systems  can be
 seen predicted from properties of Bessel
 functions. Fractional revivals with smaller
 strength are also seen.
 In the eigenbasis the expansion of the states $j$ around the
 center of the wave packet $\overline{j}$
 $E_{j}=E_{\overline{j}}+(j-\overline{j})
 {\frac {\partial E_{\overline{j}}}{\partial \overline{j}}}
 +{\frac {1}{2!}} (j-\overline{j})^{2} {\frac {\partial^{2}
 E_{\overline{j}}}{\partial^{2} \overline{j}}}
 +... = E_{\overline{j}}+  2\pi \hbar
 [-{\frac {(j-\overline{j})}{T_{cl}}}
 + {\frac {(j-\overline{j})^{2}}{2!T_{r}}} -...]$
 allows to find the time scales $T_{cl}=\hbar
 (N+1)/\sin(\overline{j}\pi/(N+1))$,
 $T_{r}=\hbar (N+1)^{2}/(\pi \cos(\overline{j}\pi/(N+1)))$, etc.
 Of course, if $|\Psi(0)\rangle$ consists of only one eigenstate with
 $j=\overline{j}$ the system shows no evolution.
 When the wave packets
 consists of many states $j$ near the
band center $E\approx 0$ then the dominant time scale is
$T_{cl}\propto \hbar (N+1)$ with $T_{r}=\infty$. When it consists
of states $j$ near the edges $E\approx \pm 2$ then $T_{cl}=\infty$
so the dominant time scale is $T_{r}\propto \hbar (N+1)^{2}$. In
Fig. 4(a) for a uniform expansion in the spatial basis we observe
period proportional to $T_{r}$  since it correspond to eigenstates
close to $E\approx \pm 2$. In Fig. 4(b) the uniform distribution
in the eigenbasis emphasises states from the band center
$E\approx 0$  which is the mean energy for this initial
superposition, so that we observe a period proportional to
$T_{cl}$.

\par
\medskip
\section{Diffusive quantum evolution}

\par
\medskip
We have also considered a chain with non-zero site potential
taking the quasi-periodic value $\epsilon_{x}=2cos(2\pi x M/N)$,
$x=1,2,...,N$ for $M$, $N$ successive Fibonacci numbers so that
the rational $M/N$ approaches the inverse golden mean ratio. In
this critical case one obtains diffusive time evolution with the
mean-square-displacement growing linearly with time $\langle
x^{2}\rangle\propto t$ which can be compared with the ballistic case
$\langle x^{2}\rangle\propto t^{2}$  studied in Chapter II. It
must be stressed again that although diffusive this is quantum
evolution and not classical.

\subsection{Space-time structures}

\par
\medskip
%
\par
 \centerline{\psfig{figure=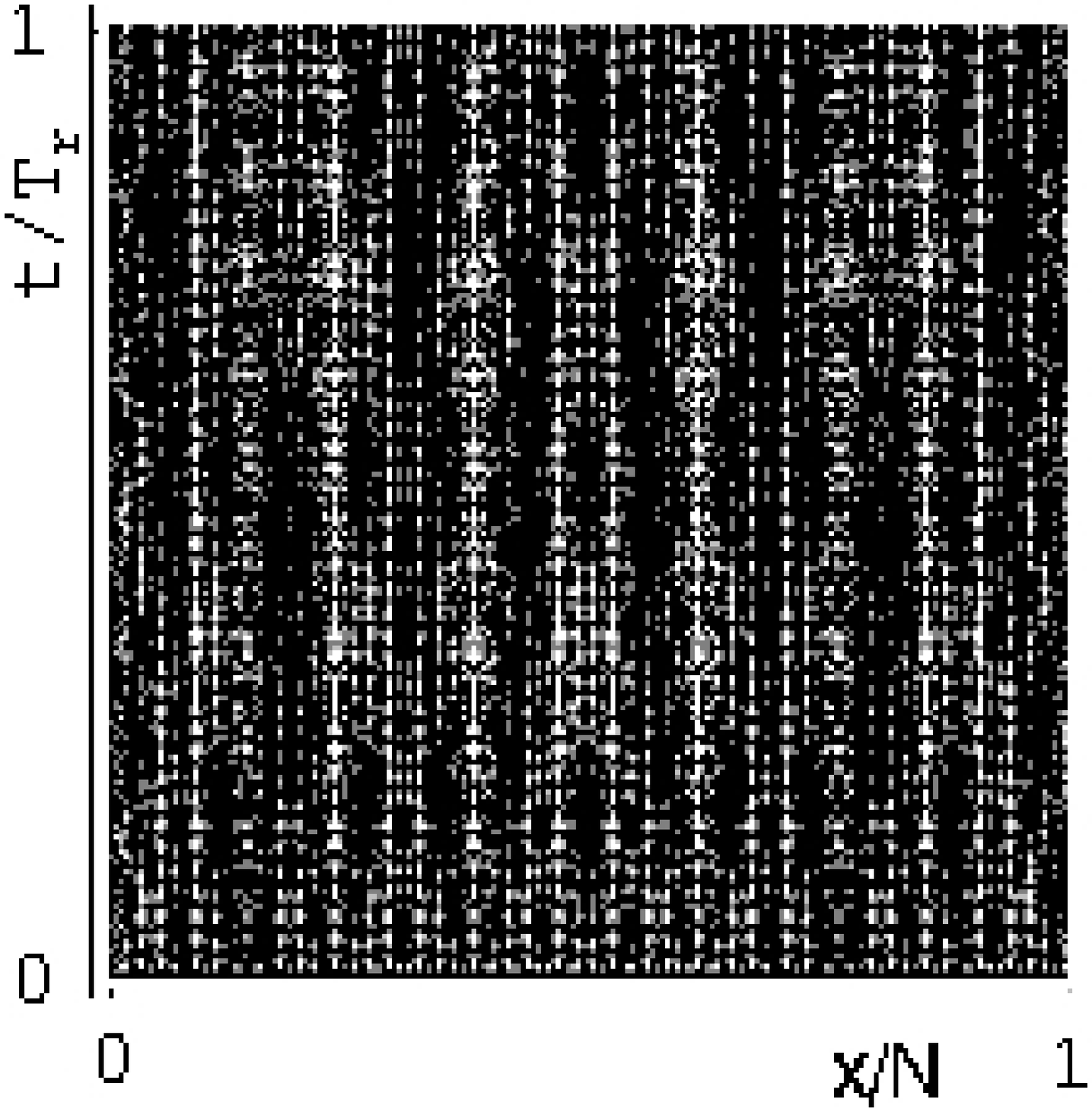,width=8.0cm}}
 \centerline{\psfig{figure=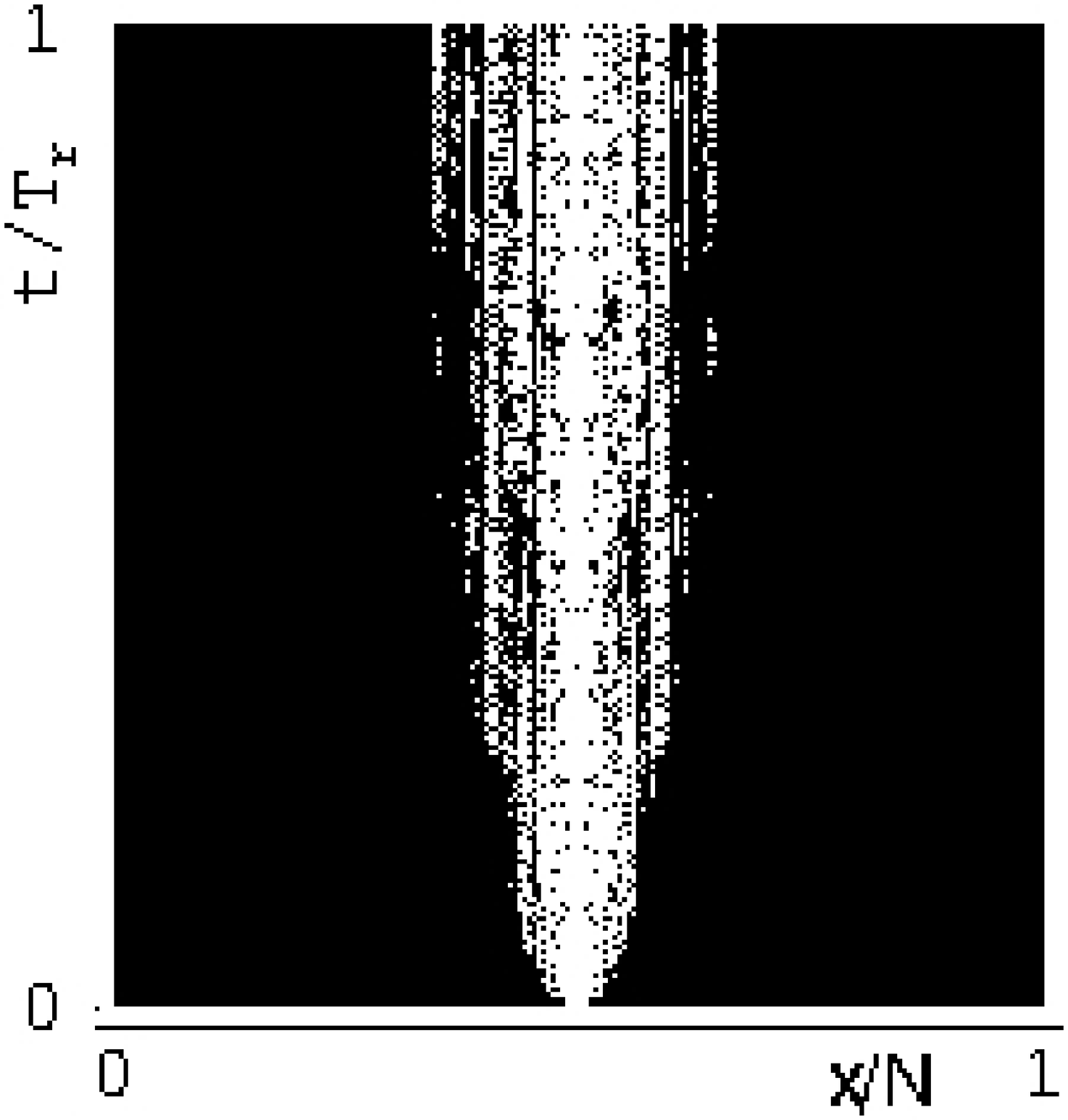,width=8.0cm}}
\par
{\footnotesize{{\bf Fig. 5. } The space-time ($x-t$)
representations for quantum diffusive evolution in a quasiperiodic
chain of length $N=987$. {\bf (a)} For an
initially uniform distributed wavepacket in space
as a function of space $x$ and time $t$. {\bf(b)} For an
initially Gaussian wave packet in space. }}
\par
\medskip

\par
\medskip
The obtained space-time structures for the diffusive case and two
spatial initial conditions are shown in Fig. 5. Again, the white
colour denotes high amplitude and the darker lower amplitude. The
figures show spectacular patterns due to quantum intereference
from scattering at the hard walls at the ends of the chains but
no quantum revivals. We show the space-time evolution for a
uniform (Fig. 5(a)) and Gaussian spatial wave packet (Fig. 5(b)).
The quantum interference gives very different results in this
case since the obtained quantum carpets display self-similarity
superimposed on the self-affinity found for ballistic evolution.
For the Gaussian in space initial state Fig. 5(b) we see slow
quantum evolution with the system not reaching the boundary
within $T_{cl}$.

\subsection{Fractal dimensions for
the probability density}

\par
\medskip
%
%
 \centerline{\psfig{figure=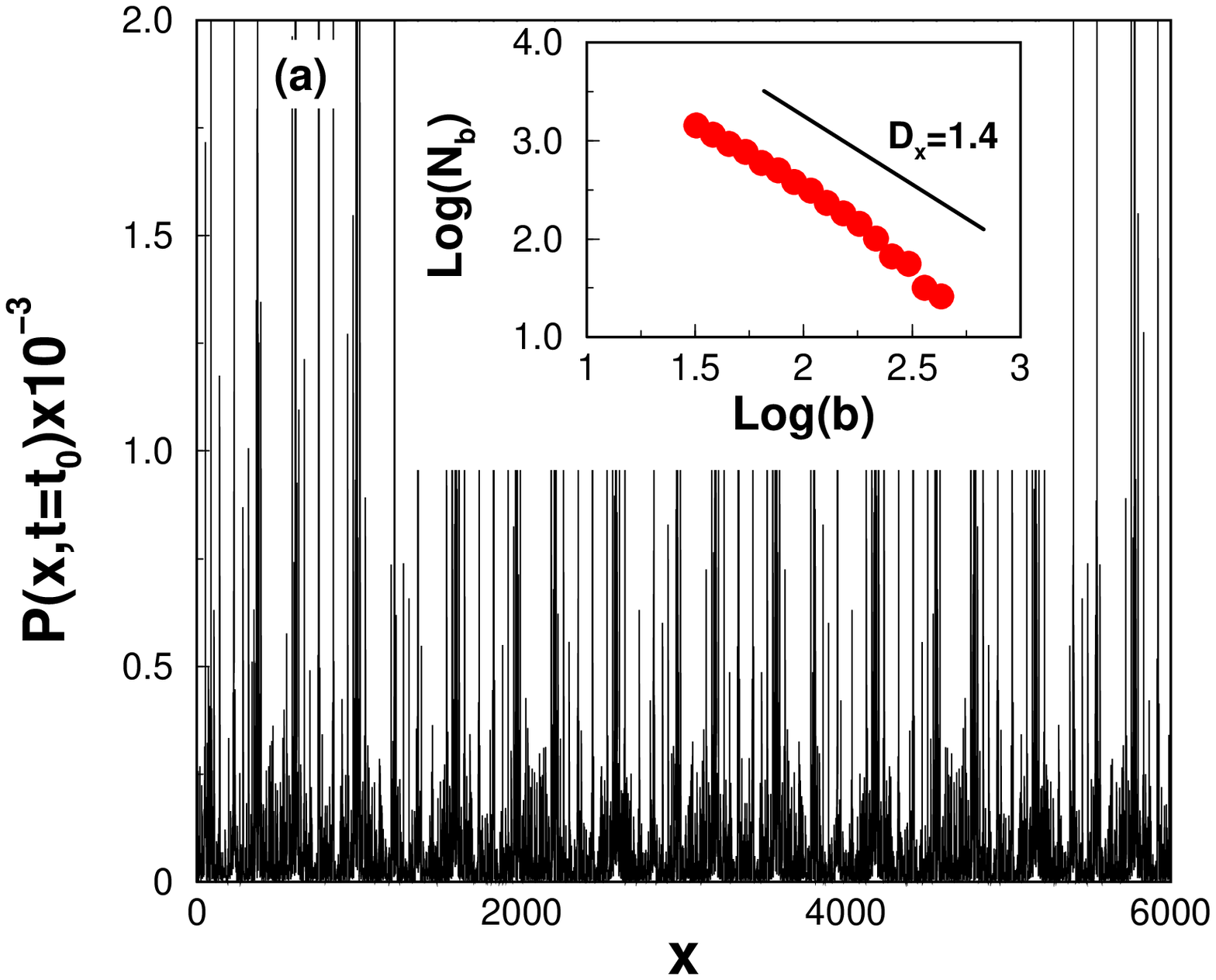,width=8.0cm}}
 \centerline{\psfig{figure=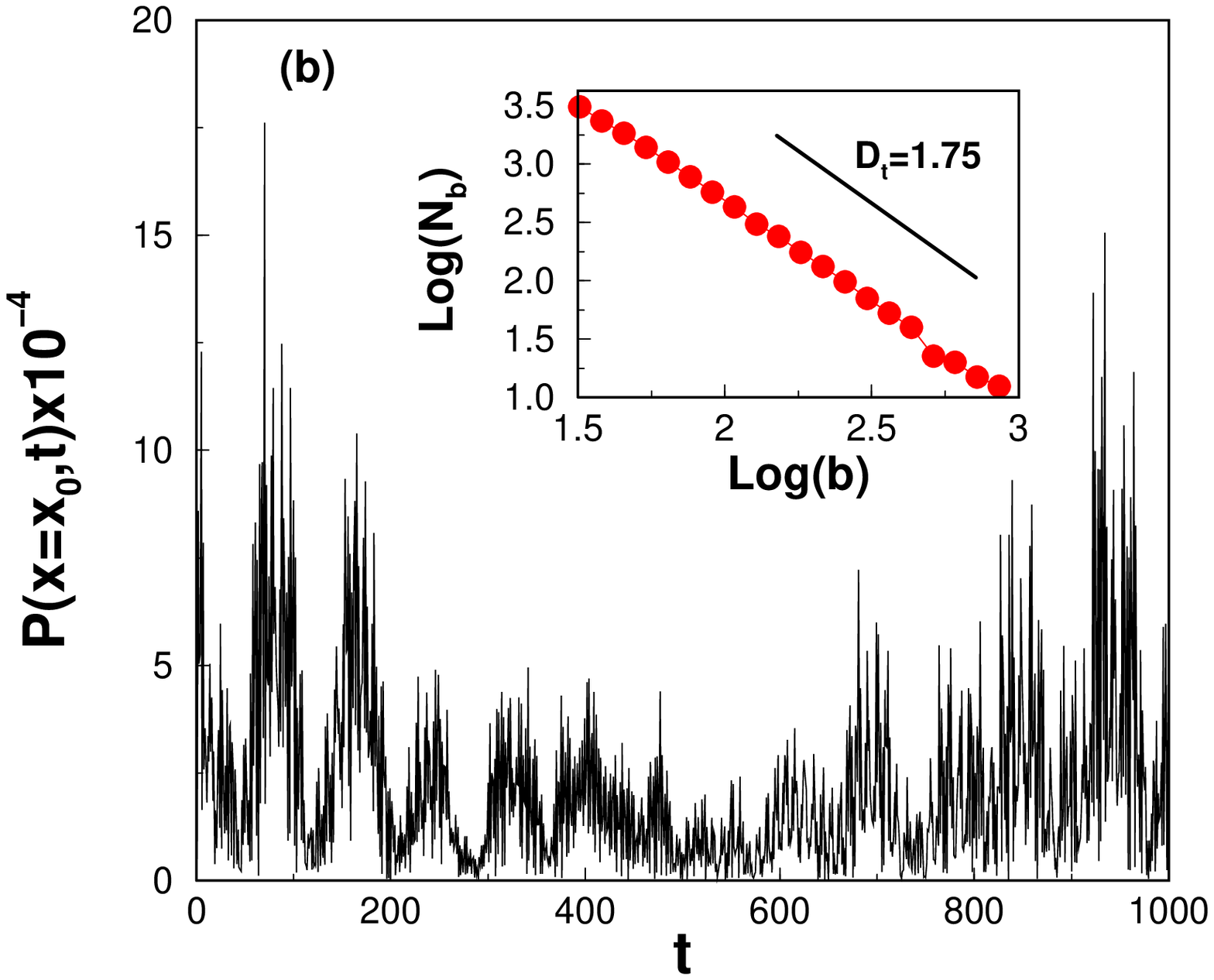,width=8.0cm}}
\par
{\footnotesize{{\bf Fig. 6. } Fractal noise for diffusion in a
chain of length $N=987$. {\bf (a)}  An initially uniform wave
packet at time $t=10$ as a function of space. {\bf (b)} As
function of time at site $x_{0}=500$.}}

\par
\medskip
We obtain similar to ballistic case self-affine fractal dimensions
displayed in Fig. (6). Apart from self-affine the
 curves are also self-similar fractals with dimensions defined in
\cite{16}. From the normalized probability measure $P(x,t)$  with
$ \sum_{x=1}^{N}P(x,t)=1$ we compute $-\sum_{x=1}^{N}P(x,t)\ln
P(x,t)$, $ \ln \sum_{x=1}^{N}P(x,t)^{2}$ and from the linear fits
vs. $\ln N$ obtain the space fractal dimensions $D_{1}^{x}$,
$D_{2}^{x}$, respectively.  Our results gave give the entropy
fractal dimension $D_{1}^{x}\approx 0.85$ for the uniform and
$\approx 0.43$ for the Gauss while for the correlation dimension
we obtain $D_{2}^{x}\approx 0.77$ for the uniform and $\approx
0.37$ for the Gauss. It must be pointed out that the fractal
dimensions $D_{1}, D_{2}$ depend strongly on initial conditions
(see Fig. 7).

\par
\medskip
%
%
%
 \centerline{\psfig{figure=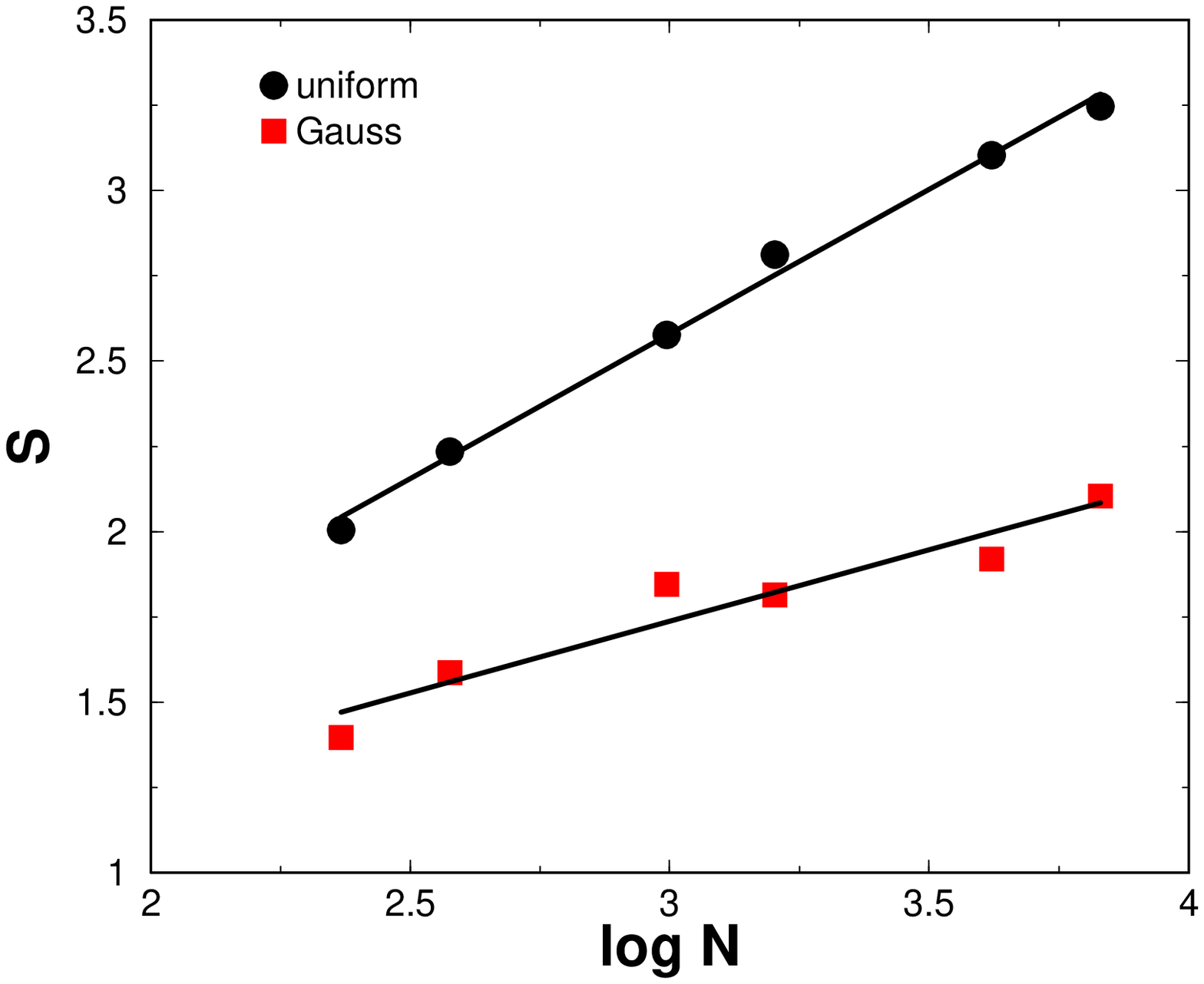,width=8.0cm}}
{\footnotesize{{\bf Fig. 7. } Entropy vs. $log N$ for the
diffusive model for an initially {\bf(a)} uniform and {\bf(b)}
gauss distribution, both in space. The information dimension
$D_{1}^{x}\approx 0.85$ for the uniform and $D_{1}^{x}\approx
0.43$ for the Gauss is obtained form the slopes of the linear
fits. Similar results exist for the correlation 
dimension $D_{2}$. }}

\par
\medskip
\subsection{``Return to the origin" probability}

\par
\medskip
The ``return to the origin" probability now shows many
fluctuations and if we define the time integrated correlation function
$C(t)=(1/t){\int_{0}^{t}} P_{0}(t')dt'$ vs. $t$. We also observe the
long-time asymptotically algebraic overall decay $P_{0}(t)\propto
{\frac {1} {\sqrt{t}}}$ known for diffusive systems.

\par
\medskip
%
%
%
\par
 \centerline{\psfig{figure=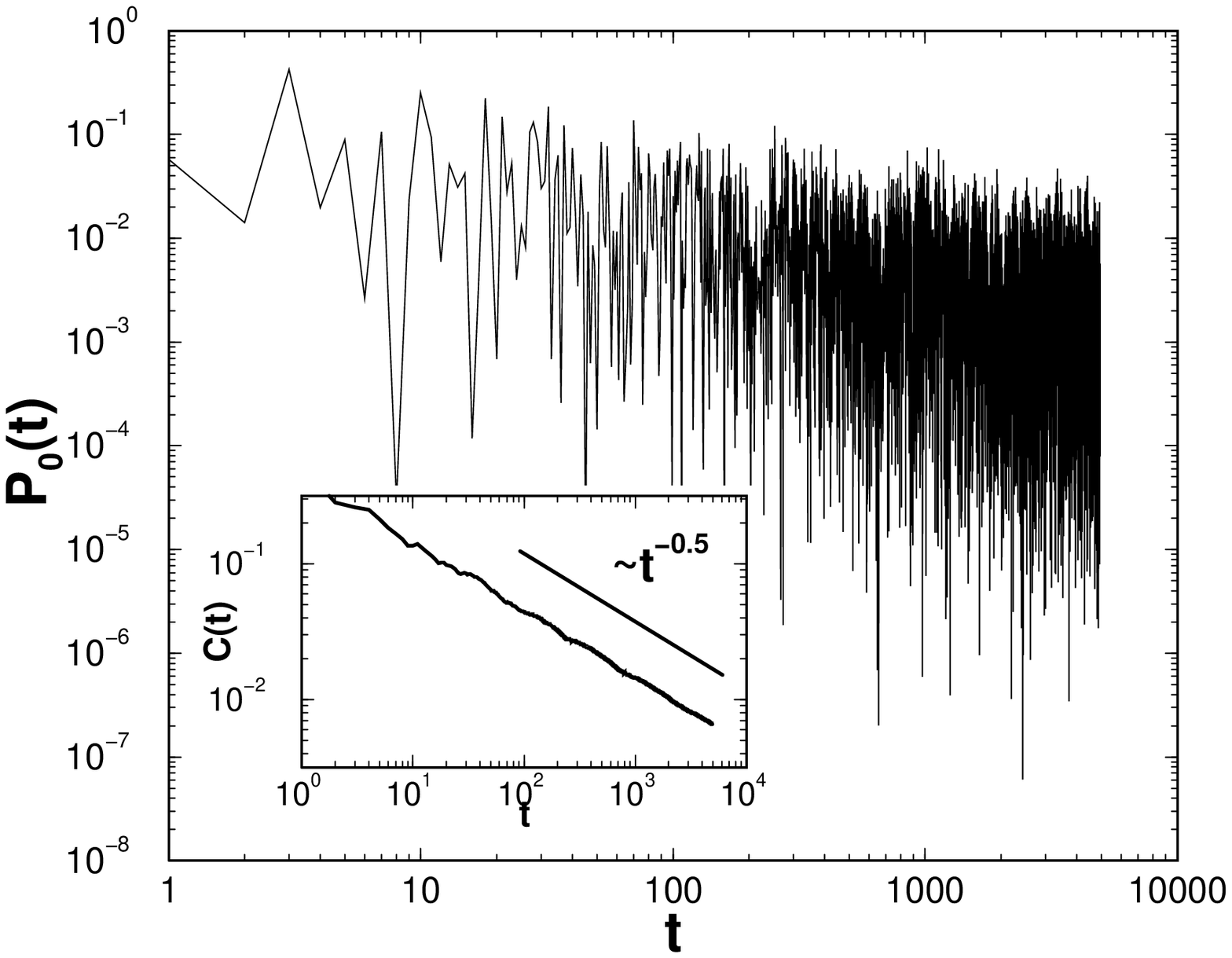,width=8.0cm}}
{\footnotesize{{\bf Fig. 8 .} The ``return to the origin
probability" for an initially uniform distributed wavepacket in a
chain of length $N=987$. The inset shows the log-log scale of the
correlation function $C(t)$ vs.  $t$ with diffusive correlation 
behavior.}}

 \medskip
\section{Discussion-Conclusions}

\par
\medskip
We studied the milder when compared to classical
quantum ballistic and diffusive evolution of wave packets.
It is different from that of classical point 
particles, nevertheless, shows fractal noise in the form
of self-affine curves.  Our calculations
are done by  discretizing the space into a tight-binding 
chain lattice. For
the ballistic case semiclassical diffusive evolution
does not exist for small $t$, to cross over to ballictic for longer times
\cite{18}, since the ballistic law is satisfied for all $t$. In this
case we display quantum memory effects by observing different
evolutions for various initial superposition wave packets
$\Psi(x,t)$. However, in agreement
with Berry's findings for a ``particle in box" a
self-affine fractal space-time $P(x,t)=|\Psi(x,t)|^{2}$
is found. For diffusive systems the evolution apart from self-affine
is also self-similar with no recurrences.

\par
\medskip
The quantum evolution could be efficiently implemented by a quantum
computer so that one can use the simulation to solve certain
computational tasks by creating more efficient algorithms. In
fact, $P_{0}(t)$ can reveal the prime factors of an encoded
number \cite{12}. On the other hand, fractals
known from non-linear physics also appear in quantum physics,
despite the fact that quantum mechanics is linear. The space
fractal dimension $D_{x}=1.5$ obtained for the spatial
distribution corresponds to Brownian $1/f^{2}$ noise familiar
from random walks. Therefore, techniques can be transferred from
non-linear physics to quantum mechanics and vice versa.
 Finally, quantum intereference can make itself obvious in scattering from
nanostructures, such as carbon nanotubes.
It can possibly influence the
temperature independent white noise known as ``shot noise" for
measurements of current in nanostructures \cite{19}.

\par
\medskip
In summary, classical systems are diffusive due to either
non-linear effects, which may lead to deterministic chaos, or
introduced randomness from outside (flipping a coin)
as for the random walk. In
these cases irreversibility naturally arises with convergence to a
Gaussian steady state distribution for any initial state. In
quantum systems the evolution is smoother, wave-like,
reversible,  with no convergence to a steady state, since quantum
memory develops. However,
we show for simple lattice systems fractal noise
due to quantum intereference from scattering at 
the boundaries. The noisy
quantum evolution seems not only independent of 
chaos but is  present for all kinds of evolution.

\end{multicols}

\end{document}